\documentclass[preprint,prd,aps,showpacs,preprintnumbers,amsmath,amssymb]{revtex4-1}
\usepackage{graphics,epsfig,subfigure}
\usepackage{diagbox}
\usepackage[usenames]{color}
\usepackage[colorlinks,
            linkcolor=black,
            anchorcolor=blue,
            citecolor=blue
            ]{hyperref}
\usepackage{color}
\usepackage{graphicx}
\usepackage{amsfonts}
\usepackage{indentfirst}
\usepackage{booktabs}

\begin{document}
\renewcommand{\baselinestretch}{1.3}
\newcommand\beq{\begin{equation}}
\newcommand\eeq{\end{equation}}
\newcommand\beqn{\begin{eqnarray}}
\newcommand\eeqn{\end{eqnarray}}
\newcommand\nn{\nonumber}
\newcommand\fc{\frac}
\newcommand\lt{\left}
\newcommand\rt{\right}
\newcommand\pt{\partial}

\title{\Large \bf Self-interacting multistate boson stars}
\author{Hong-Bo Li\footnote{lihb2017@lzu.edu.cn}, Yan-Bo Zeng\footnote{zengyb19@lzu.edu.cn}, Yan Song\footnote{songy18@lzu.edu.cn},  and Yong-Qiang Wang\footnote{yqwang@lzu.edu.cn, corresponding author
}
}

\affiliation{Research Center of Gravitation and Institute of Theoretical Physics  and  Key Laboratory for Magnetism and Magnetic of the Ministry of Education, Lanzhou University, Lanzhou 730000, China}

\begin{abstract}
In this paper, we consider rotating multistate boson stars with quartic self-interactions. In contrast to the nodeless quartic-boson stars in~\cite{Herdeiro:2015tia}, the self-interacting multistate boson stars (SIMBSs) have two types of nodes, including the $^1S^2S$ and $^1S^2P$ states. We show the  mass  $M$ of SIMBSs as a function of the synchronized frequency $\omega$, and the nonsynchronized frequency $\omega_2$ for three different cases. Moreover, for the case of two coexisting states with self-interacting potential, we study the mass $M$ of SIMBSs versus the angular momentum $J$ for the synchronized frequency $\omega$ and the nonsynchronized frequency $\omega_2$. Furthermore, for three different cases, we  analyze the coexisting phase with both the ground and  first excited states for SIMBSs.  We also calculate the maximum value of coupling parameter $\Lambda$, and find the coupling parameter $\Lambda$ exists the finite range.
\end{abstract}

\maketitle

\section{Introduction}\label{Sec1}

The scalar field dark matter (SFDM) is one of the important models of the dark matter~\cite{Sahni:1999qe,Matos:2000ng,Hu:2000ke,Padilla:2019fju}, and the model considers a spin-0 scalar field of the very small mass (typically $\sim 10^{-22} \mathrm{eV}/c^2$). Due to the ultra-light boson mass, the SFDM could form Bose-Einstein condensates (BEC) in the very early Universe, which are interpreted as the dark matter haloes. The presence of dark matter halo is inferred from its gravitational effect on a spiral galaxy rotation curve (RC). At this point boson stars (BSs) could play an important role.
The BSs were studied in the pioneering works of Kaup {\em et al.}~\cite{Kaup:1968zz,PhysRev.187.1767}, where they considered a complex scalar field coupled to Einstein gravity theory,  ones refer to this coupled system as the Einstein--Klein--Gordon~(EKG) theory.
See Refs.~\cite{Schunck:2003kk,Liebling:2012fv} for a review.

For the spherical symmetry solutions of boson stars, Deng and Huang~\cite{Deng:1998} have constructed a system consisting of  coexisting states of two scalar fields, including two ground states, and they calculated the gravitational redshift of the system. In 2010, Bernal {\em et al.}~\cite{Bernal:2009zy} have obtained the configurations with two states, a ground and first existed states, as well as they have produced the RC of multistate boson stars that are flatter at large radii than the RC of boson stars with only one state.
The study of spherical symmetry solutions of boson stars can be extended to the gauge charge~\cite{Brihaye:2004nd,Dias:2011tj,Kiczek:2020gyd} case, with de Sitter\cite{Fodor:2010hg,Cai:1997ij}, or anti-de Sitter~\cite{Buchel:2013uba,Astefanesei:2003qy} boundary conditions and the geodesics case~\cite{Eilers:2013lla}.
Moreover, the axially symmetric solutions of boson stars were studied by Schunck and Mielke~\cite{Schunck:1996,SchunM98}, Yoshida and Eriguchi~\cite{YoshiE97b},
Herdeiro and Radu~\cite{Herdeiro:2015gia}. However, they considered only ground state of the free, complex scalar field. The study of rotating axisymmetric solutions of BSs can be extended to the excited case~\cite{Wang:2018xhw}, the multistate boson stars~\cite{Li:2019mlk}.

In 1986, Colpi {\em et al.}~\cite{1986PhRvL..57.2485C} have constructed the self-interacting boson stars solutions with the type $\lambda |\psi|^4$, the sixth power $|\psi|$-terms potentials were obtained by Mielke and Scherzer~\cite{MS:81} and Kleihaus {\em et al.}~\cite{Kleihaus:2005me}. Generalizing the rotating $Q$-balls models of Volkov and W\"ohner~\cite{Volkov:2002aj} to the self-gravitating case.
In addition, in 2015,  Herdeiro and Radu have studied the quartic-BSs solutions~\cite{Herdeiro:2015tia}, which is extended to the Kerr black holes with self-interacting scalar hair.
The study of BSs can be extended to the case of charged BSs~\cite{Jetzer P:1989,Kumar:2016sxx,Kichakova:2013sza}, nonrotating and rotating axion boson stars~\cite{Guerra:2019srj,Delgado:2020udb},
Proca stars~\cite{Brito:2015pxa,Duarte:2016lig,Sanchis-Gual:2017bhw}, and in the higher-dimensional spacetime case~\cite{Hartmann:2010pm}.
The linear stability of boson stars  with respect to small oscillations was discussed by Lee and Pang in~\cite{Lee:1988av}, the study of the stability properties of boson stars extended to the quartic and sextic self-interaction term case~\cite{Kleihaus:2011sx}, dynamical $\ell$-BSs~\cite{Alcubierre:2019qnh,Jaramillo:2020rsv}, and non-linear dynamics of spinning BSs case~\cite{Sanchis-Gual:2019ljs}.
The properties of BSs have been investigated widely~\cite{Brihaye:2020klz,Valdez-Alvarado:2020vqa,Guzman:2019gqc,Kulshreshtha:2020scn,Collodel:2019uns,Kouvaris:2019nzd,Guzman:2019mat,Herdeiro:2019mbz,Choi:2019mva,Kumar:2019dbi,1801186}.
In the present work, we would like to numerically solve the  EKG equations
and give a family of rotating multistate boson stars for three different cases, including the ground state with self-interacting potential,
the first excited state with self-interacting potential and the two coexisting states with self-interacting potential.

The paper is organized as follows.
In Sect.~\ref{sec2}, we introduce  the model of
the four-dimensional Einstein gravity coupled to two self-interacting complex scalar fields $\psi_i$ $(i=1,2)$.
In Sect.~\ref{sec3}, the boundary conditions of the SIMBSs are studied.
We exhibit the numerical  results of the equations of motion and show the properties of the $^1S^2S$ and the $^1S^2P$ states for three different cases in Sect.~\ref{sec4}.
Conclusions and perspective are given in Sect.~\ref{sec5}.

\section{The model setup}\label{sec2}
We consider two massive, complex  scalar fields model coupled minimally to gravity in $3+1$ dimensions describing self-interacting multistate boson stars.
The action is
\begin{equation}
 S=\int \sqrt{-g} d^4 x \left( \frac{R}{16\pi G} + {\cal L}_{m}\right)
\end{equation}
where $R$ is the Ricci scalar, $G$ is Newton's constant and ${\cal L}_{m}$ denotes
the matter Lagrangian:
\begin{equation}
\label{lag}
 {\cal L}_{m}=-\nabla_a\psi_1^*\nabla^a\psi_1
- U(|\psi_1|)-\nabla_a\psi_2^*\nabla^a\psi_2- U(|\psi_2|)\,,
\end{equation}
\begin{equation}
  {\rm with}~~~U(|\psi_i|)= \mu_i^2\left|\psi_i\right|^2 + \lambda\left|\psi_i\right|^4  \ \ , \ \ i=1,2   \ ,
\end{equation}
where $\lambda$,  $\mu_i$,  $(i=1,2)$ are  the interaction parameter and the scalar fields mass, respectively.
It is convenient to change the dimensionless coupling parameter $\lambda$ to
\begin{equation}\label{Lambda}
  \Lambda:=  {\lambda \over 4\pi}  {M_{\rm Pl}^2 \over \mu_1^2}\,,
\end{equation}
where  $M_{Pl}^2=G^{-1}$ is the Planck mass.
For simplicity of presentation, we will set $G=c=\mu_1=1$.
The corresponding equations of motion are given by
\begin{subequations}
\begin{equation}
\frac{R_{ab}}{8\pi}=\sum_{i=1}^2 \left \{2\nabla_{(a}\psi_i^*\nabla_{b)}\psi_i+g_{ab}U(|\psi_i|)\right\},
\label{eq:EKG1}
\end{equation}
\begin{equation}
 \left(\square + \frac{\partial U}{\partial \vert\psi_i\vert^2} \right)\psi_i=0   \ \ , \ \ i=1,2   \ ,
 \label{eq:EKG2}
\end{equation}
\label{eq:EKG}%
\end{subequations}
where $\Box$ is the covariant D'Alembert differential operator.

To obtain stationary axisymmetric solutions of self-interacting multistate boson stars, we choose the ansatz as follows, see {\em e.g.}~\cite{YoshiE97b,Herdeiro:2015gia,Li:2019mlk}:
\begin{eqnarray}
 \label{ansatz}
  ds^2 = -e^{2F_0(r,\theta)}dt^2 + e^{2F_1(r,\theta)}\left( dr^2 + r^2 d\theta^2 \right) + e^{2F_2(r,\theta)}r^2\sin^2\theta \left( d\varphi - W(r,\theta)dt \right)^2 \ .
\end{eqnarray}
In addition, the ansatz of  two complex scalar fields $\psi_i$ are given by
\begin{eqnarray}
\psi_i=\phi_{i(n)}(r,\theta)e^{i(m_i\varphi-\omega_i t)},  \;\;\; n=0,1,\cdots,\;\;\; m_i=\pm1,\pm2,  \cdots, \;\;\;  i=1,2.
\label{scalar_ansatz1}
\end{eqnarray}
Here $F_0$, $F_1$, $F_2$, $W$, and  $\phi_{i(n)}$ $(i=1,2)$ are functions of the radial distance $r$ and the polar angle $\theta$, only.
Subscript  $n$  of Eq.(\ref{scalar_ansatz1}) is  named as the principal quantum number of the scalar field, and  $n=0$ is regarded as the ground state and $n\geq1$ as the excited states. Subscript $i$  are indicated by two complex  scalar fields only.
In additions, the constants $m_i$ $(i=1,2)$ and $\omega_i$ $(i=1,2)$  are indicated by the azimuthal harmonic index and the frequency of the complex scalar field, respectively.
When  $\omega_1=\omega_2=\omega$, the frequency of the scalar field is called the  synchronized frequency, while $\omega_1\neq\omega_2$ is  called the nonsynchronized frequency.
It is well known the rotating boson stars with first excited state exists two types of nodes, including radial and angular nodes. The  radial and angular nodes can seen in middle panels of Figs. 1,~4 of~\cite{Li:2019mlk}.

\section{Boundary conditions}\label{sec3}

For rotating axially symmetric boson stars, exploiting the reflection symmetry  $\theta\rightarrow\pi-\theta$ on the equatorial plane,  it is enough to consider the  range $\theta \in [0,\pi/2] $ for the angular variable.
At infinity $r\rightarrow\infty$,  the boundary conditions  are
\begin{equation}\label{rbc}
  F_0=F_1=F_2=W=\phi_{i(n)}=0,\;\;\; (i=1, 2), \;\;\;  n=0,1,\cdots,
\end{equation}
and we require the boundary conditions,
\begin{equation}\label{abc}
\partial_\theta F_0(r, 0)=\partial_\theta F_1(r, 0)=\partial_\theta F_2(r, 0) =\partial_\theta W(r, 0)=\phi_{i(n)}(r, 0)=0,\;\;\;n=0,1,\cdots,
\end{equation}
 for $\theta=0$.
For odd parity solutions, we have
\begin{equation}
\partial_\theta F_0(r,\pi/2)=\partial_\theta F_1(r,\pi/2)=\partial_\theta F_2(r,\pi/2) = \partial_\theta W(r,\pi/2)= \phi_{i(n)}(r,\pi/2) = 0,\;\;\; n=1,2,\cdots,
\end{equation}
 for $\theta=\pi/2$,
while for even parity solutions, $\partial_\theta \phi_{i(n)}(r, \pi/2) = 0$ with $ n=1,2,\cdots$.

At the origin we require 
\begin{eqnarray}
 \phi_{i(n)}(0, \theta) = 0,  \nonumber\\
 \partial_r W(0, \theta) = 0.
\end{eqnarray}
We note that the values of $F_0(0, \theta), F_1(0, \theta), F_2(0, \theta)$, and  $W(0,\theta)$ are the constants that are not dependent of the polar angle $\theta$.

Near the boundary $r\rightarrow\infty$, on the other hand, the mass of boson stars $M$  and total angular momentum  $J$  are extracted from the  asymptotic behavior of  the metric functions,
\begin{eqnarray}
\label{asym}
g_{tt}= -1+\frac{2GM}{r}+\cdots, \nonumber\\
g_{\varphi t}= -\frac{2GJ}{r}\sin^2\theta+ \cdots.
\end{eqnarray}
\section{Numerical results}\label{sec4}

In this section, we will solve  the above coupled Eqs. (\ref{eq:EKG1}) and (\ref{eq:EKG2}) with the ansatzs (\ref{ansatz}) and (\ref{scalar_ansatz1}) numerically.
It is convenient to introduce a new coordinate $ x \equiv\frac{r}{1+r}$,
which implies that the new radial coordinate $x  \in [0,1]$.  Thus, the inner and outer boundaries of the shell
are fixed at $x = 0$ and  $x = 1$, respectively.  By exploiting the reflection symmetry  $\theta\rightarrow\pi-\theta$ on the equatorial plane,  it is enough to consider the  range $\theta \in [0,\pi/2] $ for the angular variable.
All numerical calculations are based on the finite element methods.
Typical grids used have sizes of $100\times100$ in the integration region $0\leq x\leq 1$ and $0\leq\theta\leq\frac{\pi}{2}$. Our iterative process is the Newton-Raphson method, and the relative error for the numerical solutions in this work is estimated to be below $10^{-5}$.

Next,  we will discuss the SIMBSs, including the principal quantum number $n=0$, which is the ground state case and
the principal quantum number $n=1$, which belongs to the case of the first excited state. Besides, we exhibit two classes of  radial $n_r=1$  and angular  $n_\theta=1$ node solutions, respectively.
As noted above, for the case of  rotating boson stars with first excited state, there is a similar situation in atomic theory and quantum mechanics, the first excited state of hydrogen has an electron in the 2s orbital and 2p orbital, which correspond to the radial and angle node, respectively. Therefore, the coexisting states of two  scalar fields, which have a ground state and a first excited state with a radial node $n_r=1$, is  named  as the $^1S^2S$ state as well as the  coexistence of  a ground state and a first excited state with an angle node $n_\theta=1$ is called the $^1S^2P$ state.

\subsection{ Case: The ground state with self-interacting potential }
For $U(|\psi_1|)\neq0$, $U(|\psi_2|)=0$, the matter Lagrangian ${\cal L}_{m}^{(1)}$ is given by
\begin{equation}
\label{lag}
 {\cal L}_{m}^{(1)}=-\nabla_a\psi_1^*\nabla^a\psi_1- U(|\psi_1|)-\nabla_a\psi_2^*\nabla^a\psi_2-\mu_2^2|\psi_2|^2\,,
\end{equation}
${\rm with}~U(|\psi_1|)= \mu_1^2\left|\psi_1\right|^2 + \lambda_1\left|\psi_1\right|^4$.
This type of solutions denotes the ground state with self-interacting potential.
We use the same relation between the parameters $\Lambda_1$ and $\lambda_1$ given in Eq.(\ref{Lambda}).
\subsubsection{$^1S^2S$ state}
\begin{figure}[h!]
\begin{center}
\includegraphics[height=.275\textheight]{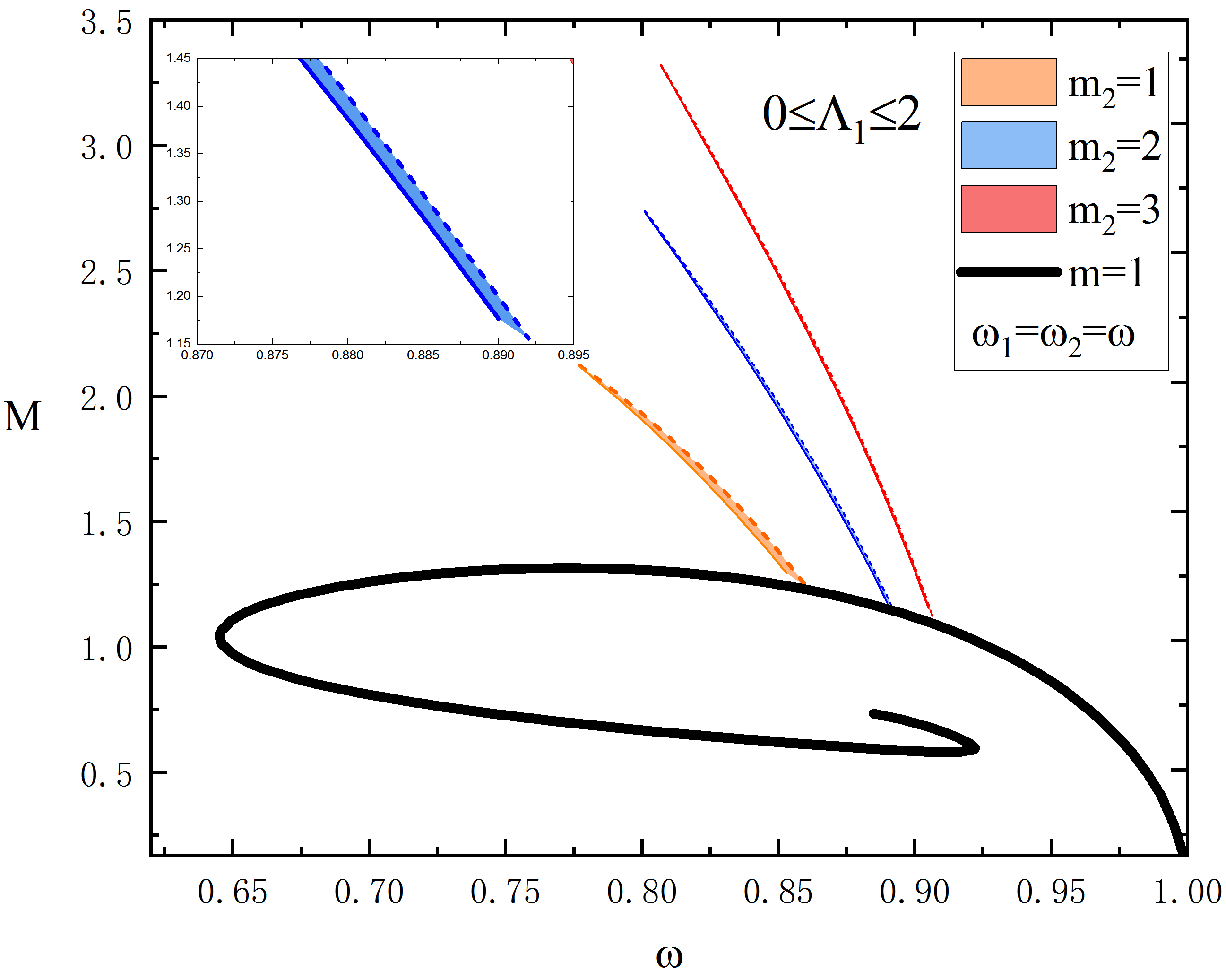}
\includegraphics[height=.28\textheight]{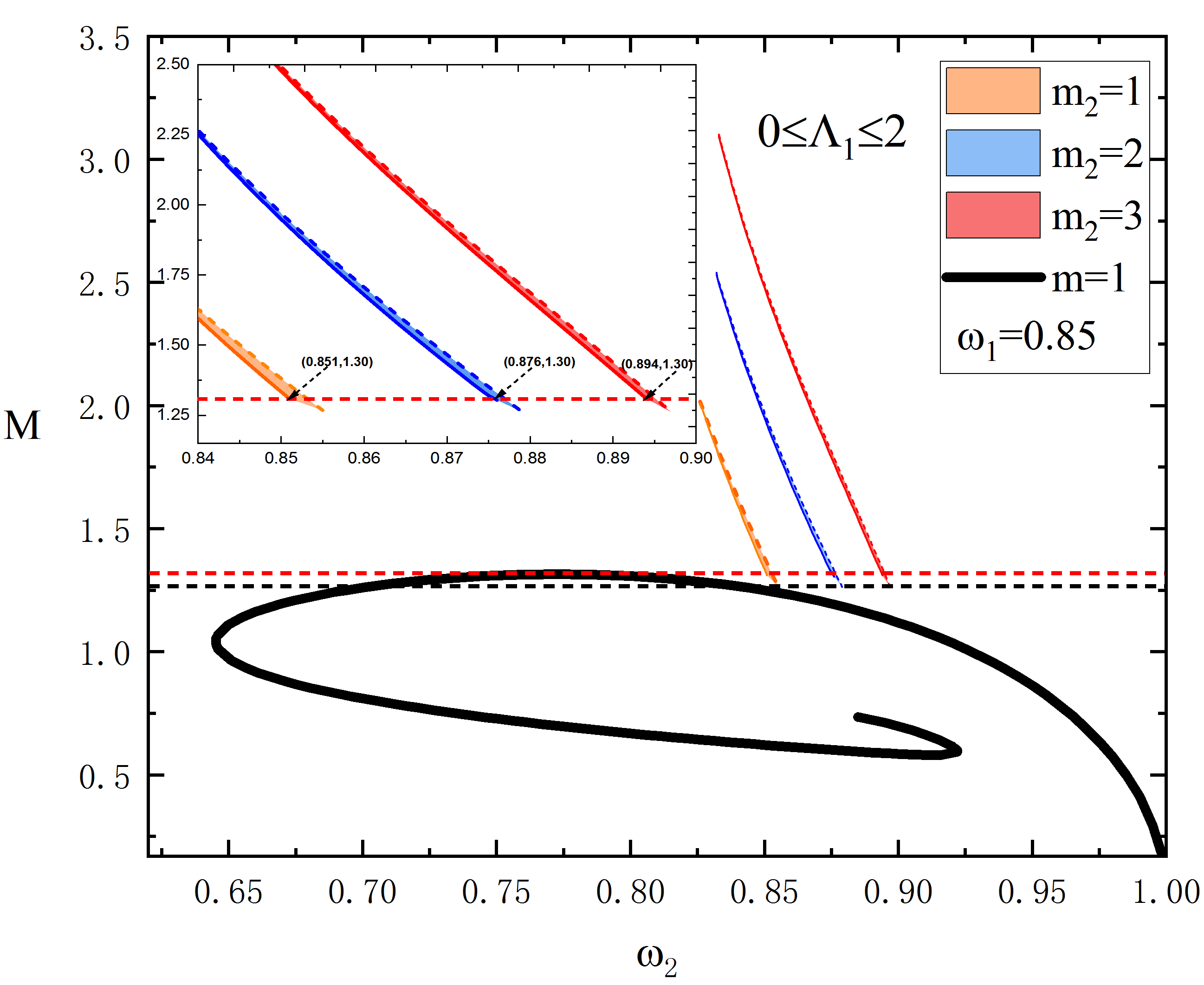}
\end{center}
\caption{\textit{Left}: The mass $M$ of the SIMBSs as a function of the synchronized frequency $\omega$ with the azimuthal harmonic indexes $m_{2}=1~,2,~3$. \textit{Right}: The mass $M$ of the  SIMBSs as a function of the nonsynchronized frequency $\omega_2$ with $\omega_1=0.85$. The horizon red dashed line indicates the mass $M=1.30$, and the black dashed line indicates the mass $M=1.26$. In both panels, the dashed, solid lines denote $\Lambda_1=0$ and $\Lambda_1=2$, respectively.
The black spiral curve indicates the ground state with an azimuthal harmonic index $m=1$.  All solutions have $\mu_1=1,\mu_2=0.93$, and $m_1=1$.}
\label{ground-1s2s}
\end{figure}

In this subsection, we will analyze the solutions with an even-parity scalar field. The value of the scalar field $\phi_i$ $(i=1,2)$ can be seen in of Fig. 1 of~\cite{Li:2019mlk}.
According to the numerical results, we find that there exists a maximal mass of multistate boson stars for parameter $\Lambda_1<\Lambda_c=2.8$.
To study the properties of the SIMBSs, we exhibit in Fig.~\ref{ground-1s2s} the mass $M$
of self-interacting multistate boson stars versus the synchronized frequency $\omega$ and the nonsynchronized frequency $\omega_2$ with the azimuthal harmonic index $m_{2}=1~,2,~3$ for fixed values of the coupling constant $\Lambda_1$.

In the left panel of Fig.~\ref{ground-1s2s}, we exhibit the mass $M$ versus the synchronized frequency $\omega$ with the azimuthal harmonic index $m_{2}=1~,2,~3$, denoted by the origin, blue and red lines, respectively, and  the shaded regions indicate $0\leq\Lambda_1\leq2$.  More details are given in the inset plotted in the left panel of Fig.~\ref{ground-1s2s}.
We observe that for the same synchronized frequency and  coupling parameters, the free, multistate boson stars possess a higher mass than the SIMBSs.
Moreover, we observe that for the fixed coupling parameter $\Lambda_1=0$ and $\Lambda_1=2$,  the multistate boson stars exist only a stable branch, which is different from  the quartic-BSs case in Ref.~\cite{Herdeiro:2015tia}.
In the right panel of Fig.~\ref{ground-1s2s}, we show the mass $M$ versus the nonsynchronized frequency $\omega_2$ for the fixed value of $\omega_1=0.85$, as well as the shaded regions indicate $0\leq\Lambda_1\leq2$. We can see that the free, multistate boson stars presents a higher value for the mass,  which is different from the quartic-BSs in Ref.~\cite{Herdeiro:2015tia}. From the inset plotted of the right of Fig.~\ref{ground-1s2s},
we found that the minimum value of mass of the SIMBSs is the constant value $M=1.30$, three coordinates correspond to $(0.851,~1.30)$,  $(0.873,~1.30)$, and $(0.894,~1.30)$, respectively.
Table~\ref{table1} shows for the different values of $\Lambda_1$, the  domain of existence of the  mass $\mu_2$. From the Table~\ref{table1}, it is apparent that as the coupling parameter $\Lambda_1$ increases, the domain of existence of $\mu_2$ decreases.
Here we note that, the SIMBSs case does not occur with another branch.
Moreover, we find the another branch do not depend on the  values of $\lambda$, in order to verify whether there exists another family of multistate solutions,  we also use the same method in~\cite{Li:2019mlk} to seek for  the new family of SIMBSs case (The more detailed analysis process can be seen from the sixth  paragraph of section IV (A) and in Fig. 3 of~\cite{Li:2019mlk}).
\begin{table}[!htbp]
		\centering
		\begin{tabular}{|c|c|c|c|c|}
			\hline
\diagbox{$m_2$}{$\mu_{2}$}{$\Lambda_{1~~}$}&$0$&$1$&$2$&$2.5$\\
            \hline
			$1$&$0.924\sim0.959$&$0.926\sim0.959$&$0.929\sim0.959$&$0.930\sim0.959$\\
			\hline
		\end{tabular}
\caption{The domain of existence of the  mass $\mu_2$ of the scalar field $\phi_2$ with various values of the coupling parameter $\Lambda_1$. All solutions have $\mu_1=1$, $m_1=1$, and $m_2=1$.}
\label{table1}
	\end{table}
\subsubsection{$^1S^2P$ state}
In order to compare the results with even-parity scalar field,
in this subsection we also study the solutions with odd-parity scalar field mentioned in~\cite{Herdeiro:2015tia,Li:2019mlk,Wang:2018xhw}.
Meanwhile, in the middle panels of Fig.~5 of~\cite{Li:2019mlk}, we can see that the odd function $\phi_2$ is antisymmetric with respect to the x-axis for the angular variable $0\leq\theta\leq\frac{\pi}{2}$. Besides, according to the numerical results, we find no solutions of self-interacting multistate boson stars for  $\Lambda_1\geq\Lambda_c=5.8$.
\begin{figure}[h!]
\begin{center}
\includegraphics[height=.25\textheight,width=.34\textheight, angle =0]{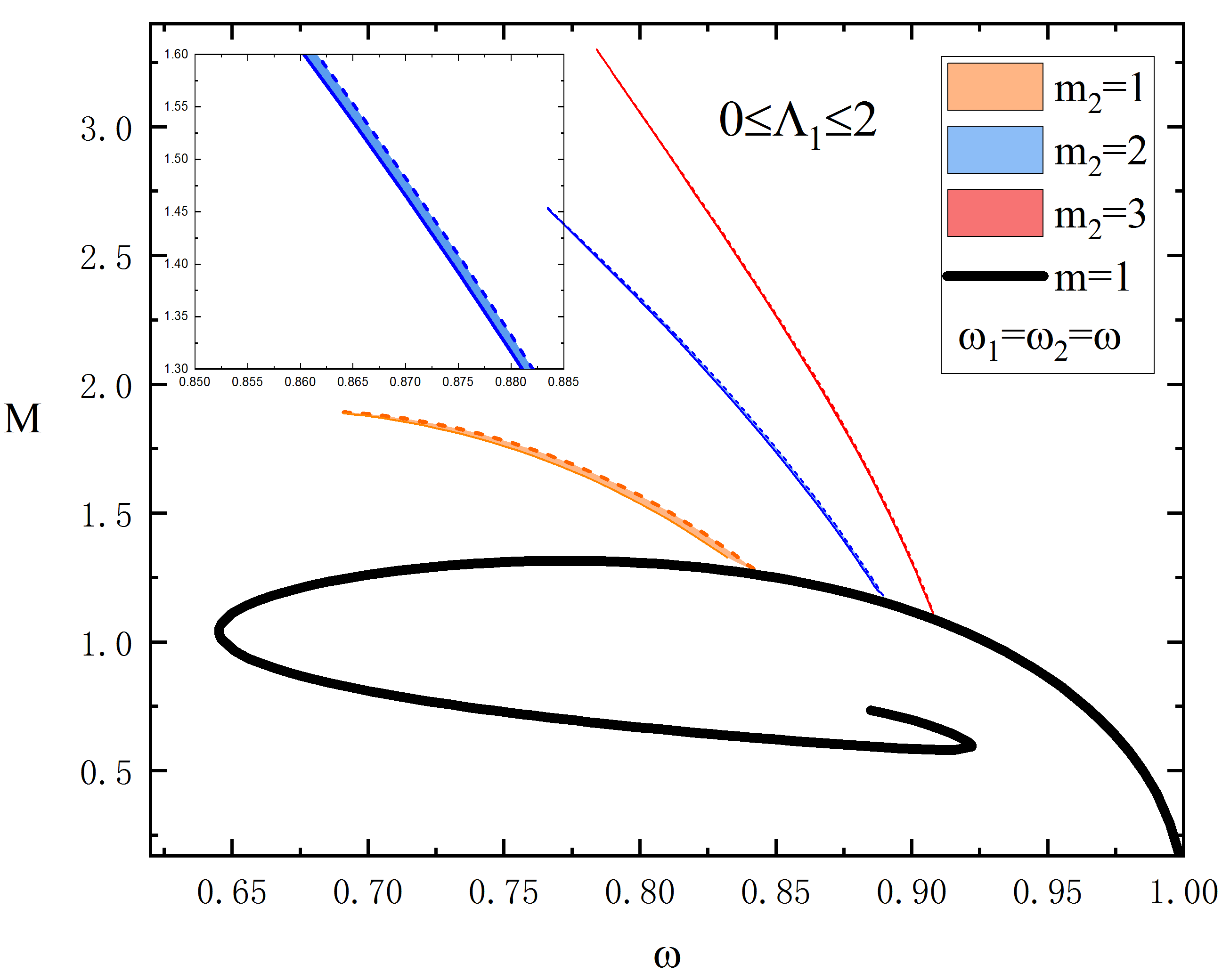}
\includegraphics[height=.253\textheight,width=.34\textheight, angle =0]{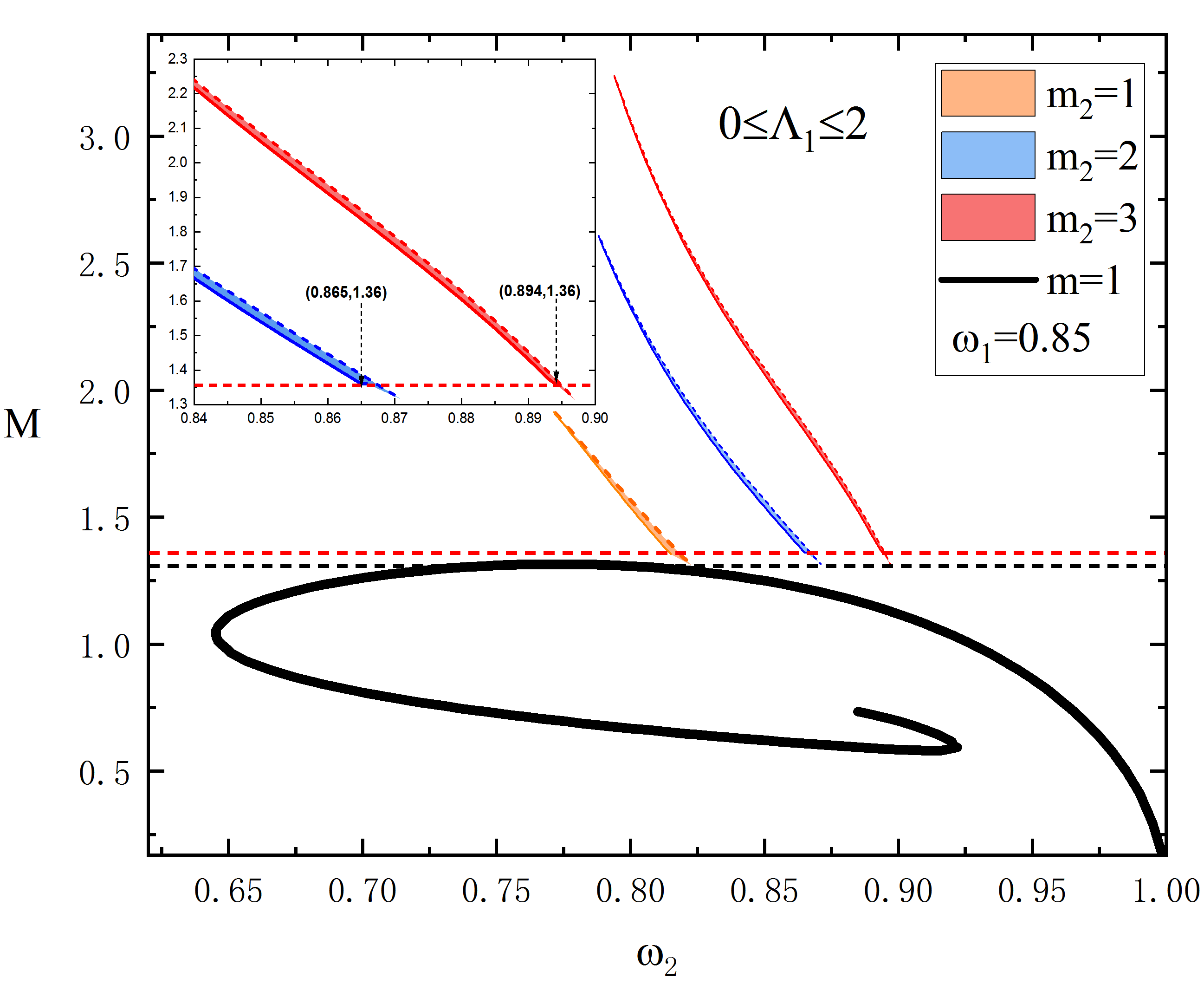}
\end{center}
\caption{\textit{Left}: The mass $M$ of the SIMBSs versus the synchronized frequency $\omega$ with the azimuthal harmonic indexes $m_{2}=1~,2,~3$. \textit{Right}: The mass $M$ of the  SIMBSs versus the nonsynchronized frequency $\omega_2$ with $\omega_1=0.85$. The horizon red dashed line indicates the mass $M=1.36$,  and the black dashed line indicates the mass $M=1.31$. In both panels, the dashed, solid lines denote $\Lambda_1=0$ and $\Lambda_1=2$, respectively.
The black spiral curve indicates the ground state with an azimuthal harmonic index $m=1$. All solutions have $\mu_1=1,\mu_2=0.93$, and $m_1=1$.}
\label{ground-1s2p}
\end{figure}

After obtaining the numerical solutions of SIMBSs, in the left panel of  Fig.~\ref{ground-1s2p}, we exhibit the mass $M$ of the SIMBSs versus the frequency $\omega$ with the azimuthal harmonic index $m_{2}=1$ (origin shaded area), $m_{2}=2$  (blue shaded area), $m_{2}=3$ (red shaded area), respectively. In the inset panel, we show the detail of the blue shaded area with  $m_{2}=2$.
We observe that the SIMBSs  exhibits only a stable branch with $m_{2}=1~,2,~3$, which is similar to the case of free, multistate boson stars in Ref.~\cite{Li:2019mlk}.
We again observe that with the decrease of the frequency $\omega$, the mass of SIMBSs keeps increasing. However, we can see that the $^1S^2P$ state curves exhibit only a stable branch with $m_{2}=1~,2,~3$, which is similar to the case of free, multistate boson stars in Ref.~\cite{Li:2019mlk}, as well as the mass of the SIMBSs is smaller than the case of free, multistate boson stars.

In the right panel of Fig.~\ref{ground-1s2p}, we plot the mass of the SIMBSs versus the nonsynchronized frequency $\omega_2$ for the $\omega_1=0.85$.
We observe that the free, multistate boson stars have higher mass level than the SIMBSs.
In Table \ref{table2}, we present the  domain of existence of the  mass $\mu_2$ of the scalar field $\phi_2$ for the different values of $\Lambda_1$. We note that with the increase of coupling parameter $\Lambda_1$, the domain of existence of the $\mu_2$ begins to decrease.
\begin{table}[!htbp]
		\centering
		\begin{tabular}{|c|c|c|c|c|}
			\hline
\diagbox{$m_2$}{$\mu_{2}$}{$\Lambda_{1~~}$}&$0$&$1$&$2$&$2.5$\\
            \hline
			$1$&$0.898\sim0.967$&$0.904\sim0.967$&$0.909\sim0.967$&$0.912\sim0.967$\\
			\hline
		\end{tabular}
\caption{The domain of existence of the  mass $\mu_2$ of the scalar field $\phi_2$ with various values of the coupling parameter $\Lambda_1$. All solutions have $\mu_1=1$, $m_1=1$, and $m_2=1$.}
\label{table2}
	\end{table}
\subsection{Case: The first excited state with self-interacting potential}
In order to obtain the solutions of first excited state with self-interacting potential, we choose the self-interacting potential $U(|\psi_1|)=0$, $U(|\psi_2|)\neq0$.
In this case, the matter Lagrangian ${\cal L}_{m}^{(2)}$ is given by
\begin{equation}
\label{lag}
 {\cal L}_{m}^{(2)}=-\nabla_a\psi_1^*\nabla^a\psi_1-\mu_1^2|\psi_1|^2-\nabla_a\psi_2^*\nabla^a\psi_2- U(|\psi_2|)\,,
\end{equation}
${\rm with}~U(|\psi_2|)= \mu_2^2\left|\psi_2\right|^2 + \lambda_2\left|\psi_2\right|^4$.
We use the same relation between the parameters $\Lambda_2$ and $\lambda_2$ given in Eq.(\ref{Lambda}).

\subsubsection{$^1S^2S$ state}
\begin{figure}[h!]
\begin{center}
\includegraphics[height=.28\textheight]{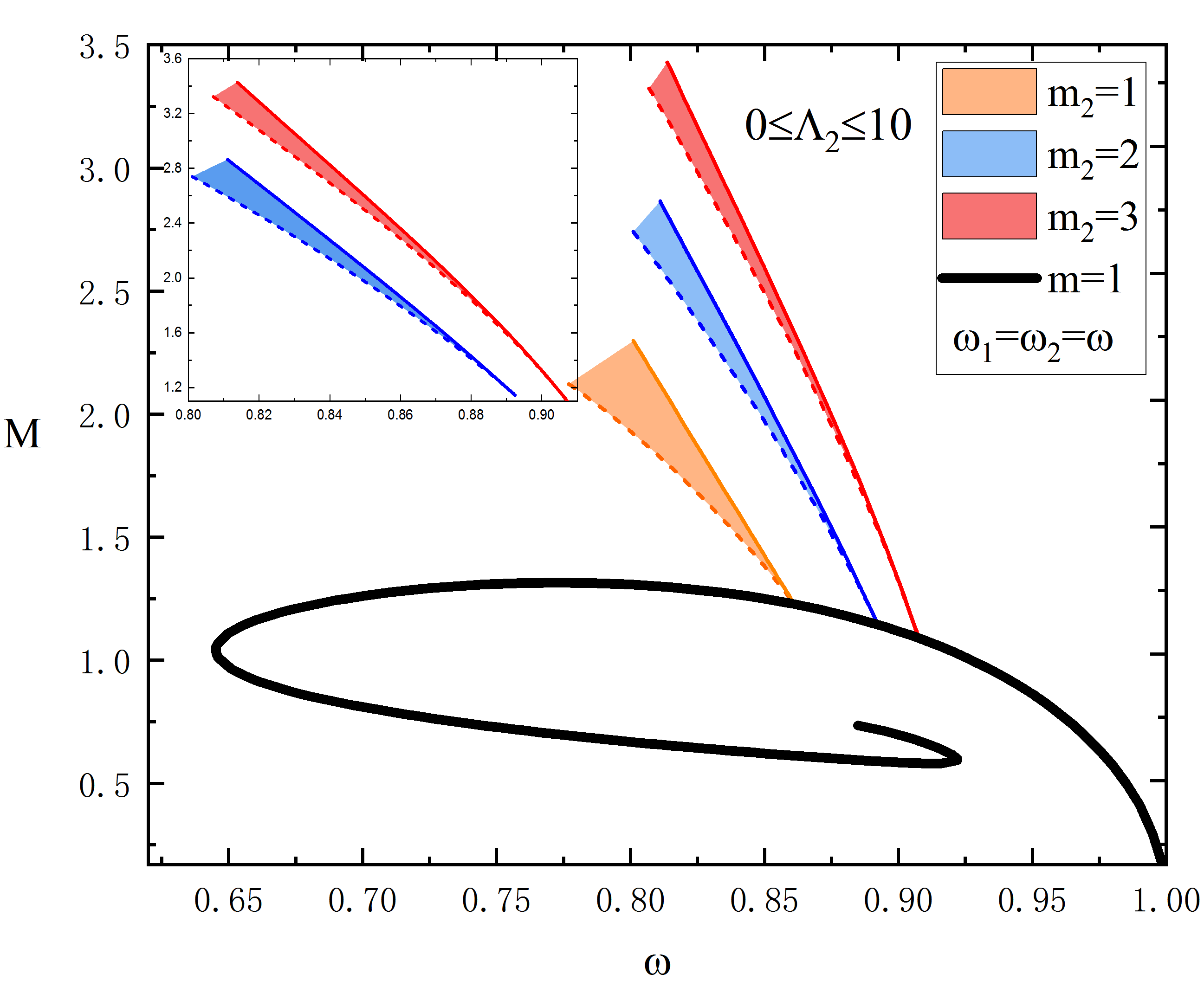}
\includegraphics[height=.28\textheight]{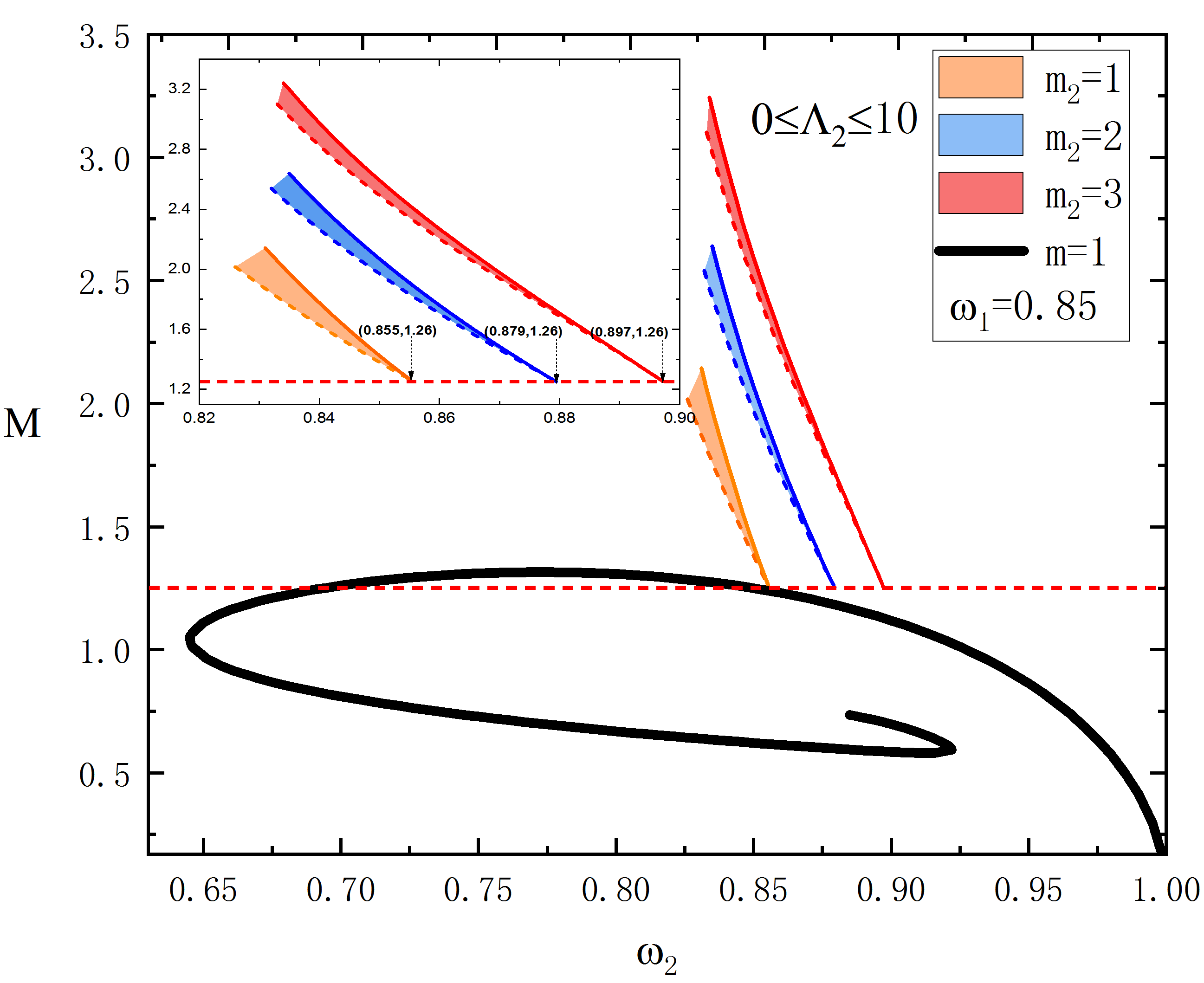}
\end{center}
\caption{\textit{Left}: The mass $M$ of the SIMBSs as a function of the synchronized frequency $\omega$ with the azimuthal harmonic indexes $m_{2}=1~,2,~3$. \textit{Right}: The mass $M$ of the  SIMBSs as a function of the nonsynchronized frequency $\omega_2$ with $\omega_1=0.85$. The horizon red dashed line indicates the mass $M=1.26$. In both panels, the dashed and solid lines stand for $\Lambda_2=0$ and $\Lambda_2=10$, respectively. The black spiral curve indicates the ground state with $m=1$ and all solutions have $\mu_1=1,\mu_2=0.93$, $m_1=1$.}
\label{excited-1s2s}
\end{figure}

According to the numerical results, we find that the  range of coupling parameter $\Lambda_2\in(0,388)$.
In the left panel of Fig.~\ref{excited-1s2s}, we show the mass $M$ of the SIMBSs versus the synchronized frequency $\omega$ with the azimuthal harmonic index $m_{2}=1~,2,~3$, represented by the origin, blue, and red lines, respectively, and  the shaded regions indicate $0\leq\Lambda_2\leq10$.  More details are given in the inset plotted in the left panel of Fig.~\ref{excited-1s2s}.
We observe that for the fixed coupling parameters $\Lambda_2=0$ and $\Lambda_2=10$,  all the  multistate boson stars curves exist a unique branch, which is different from the case of ground state in Ref.~\cite{Herdeiro:2015tia}. Moreover, we observe that for the same synchronized frequency and coupling parameters, the self-interacting multistate boson stars solutions is heavier than the case of free, multistate boson stars.
This result can be understood that when the synchronized frequency $\omega$ tends to its maximum, the scalar field of the first excited state could reduce to zero and there exists only a single scalar field of the ground state.  So, three sets of the multistate boson stars  intersect with the ground state solutions, respectively.

In the right panel of Fig.~\ref{excited-1s2s}, we present the mass $M$ of the SIMBSs versus the non-synchronized frequency $\omega_2$ for the fixed value of $\omega_1=0.85$, and the dashed lines and solid lines stand for $\Lambda_2=0$ and $\Lambda_2=10$, respectively. From the graphic, it is obvious that the SIMBSs has higher mass,  which is similar to the case of quartic-BSs in Ref.~\cite{Herdeiro:2015tia}. In the inset plotted of the right of Fig.~\ref{excited-1s2s},
we found that the minimum value of mass of the SIMBSs is the constant value $M=1.26$, which is the same as free, multistate boson stars solutions.
This means that  with the  nonsynchronized frequency $\omega_2$ increases,
the first excited state could decrease to zero and the mass of the multistate boson stars is provided by the ground state. While,  we fixed the value of $\omega_1=0.85$, the minimal mass of the multistate boson stars is always a constant value $M=1.26$ for the different azimuthal harmonic indexes $m_2=2,~3$.
Hence, three coordinates correspond to $(0.855,~1.26)$,  $(0.879,~1.26)$, and $(0.897,~1.26)$, respectively.
To explore the influence of the different typical coupling parameters, we exhibit for various values of $\Lambda_2$, the domain of existence of the  mass $\mu_2$ of the scalar field $\phi_2$ is shown in Table \ref{table3}. It is observe that, the domain of existence of $\mu_2$ decreases with increasing coupling parameter $\Lambda_2$.
\begin{table}[!htbp]
		\centering
		\begin{tabular}{|c|c|c|c|c|}
			\hline
\diagbox{$m_2$}{$\mu_{2}$}{$\Lambda_{2~~~}$}&$0$&$10$&$20$&$35$\\
            \hline
			$1$&$0.924\sim0.959$&$0.924\sim0.953$&$0.924\sim0.949$&$0.924\sim0.945$\\
			\hline
		\end{tabular}
\caption{The domain of existence of the  mass $\mu_2$ of the scalar field $\phi_2$ with various values of the coupling parameter $\Lambda_2$. All solutions have $\mu_1=1$, $m_1=1$, and $m_2=1$.}
\label{table3}
	\end{table}

\subsubsection{$^1S^2P$ state}
\begin{figure}[h!]
\begin{center}
\includegraphics[height=.27\textheight,width=.34\textheight, angle =0]{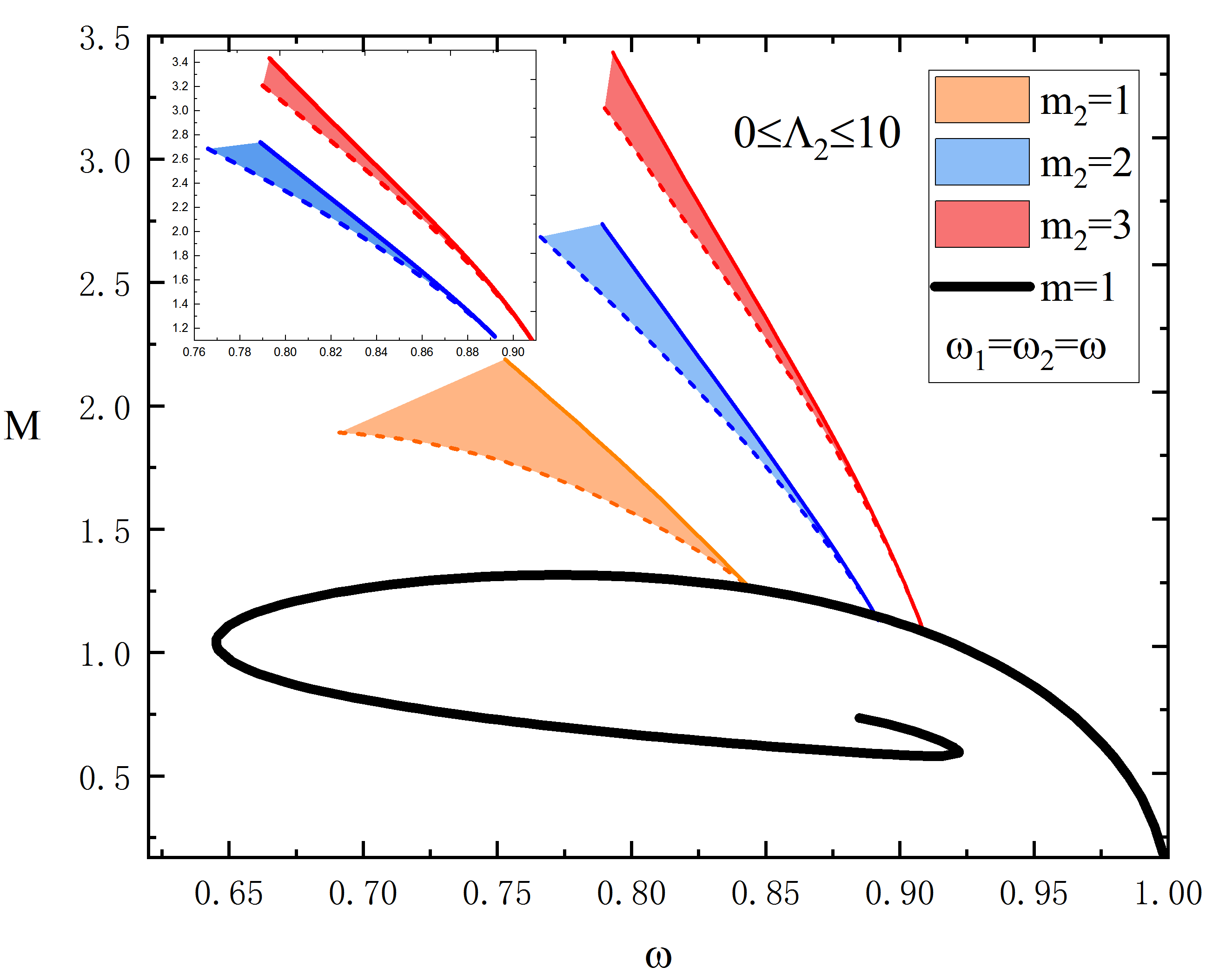}
\includegraphics[height=.27\textheight,width=.34\textheight, angle =0]{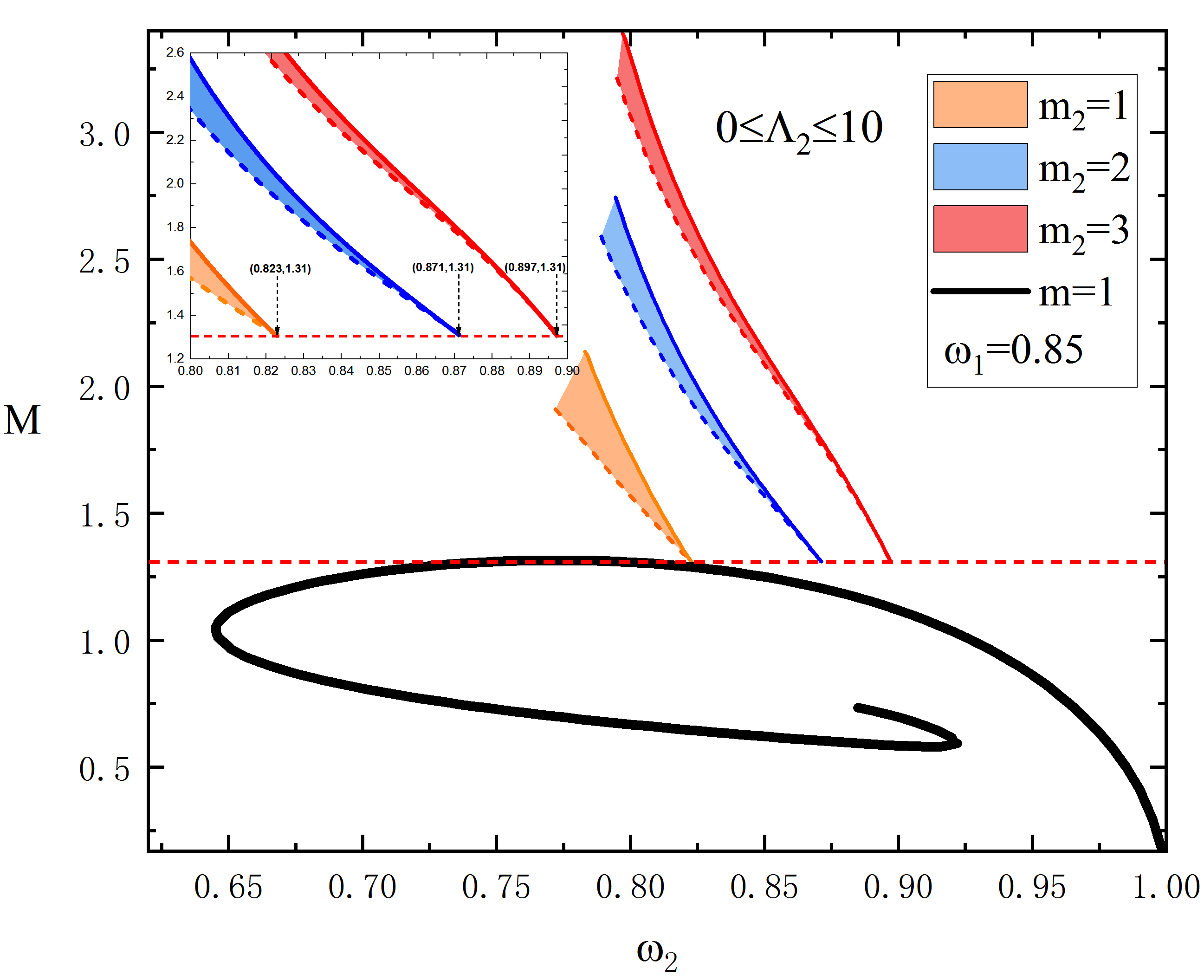}
\end{center}
\caption{\textit{Left}: The mass $M$ of the  SIMBSs as a function of the synchronized frequency $\omega$ with the azimuthal harmonic indexes $m_{2}=1~,2,~3$. \textit{Right}: The mass $M$ of the  SIMBSs as a function of the nonsynchronized frequency $\omega_2$ with $\omega_1=0.85$. The horizon red dashed line indicates the mass $M=1.31$. In both panels, the dashed and solid lines stand for $\Lambda_2=0$ and $\Lambda_2=10$, respectively. The black spiral curve indicates the ground state with $m=1$ and all solutions have $\mu_1=1,\mu_2=0.93$, $m_1=1$.}
\label{excited-1s2p}
\end{figure}

As typical examples for our numerical results,
we obtain the range of coupling parameter $\Lambda_2\in(0,41)$. Meanwhile,
we show in Fig.~\ref{excited-1s2p} the SIMBSs mass $M$ as  functions of the synchronized frequency $\omega$ and  nonsynchronized frequency $\omega_2$, respectively, by taking  $m_{2}=1$ (origin shaded region), $m_{2}=2$ (blue shaded region), $m_{2}=3$ (red shaded region). In the left panel, we observe that as azimuthal harmonic index $m_{2}$ increases, the maximum value of the synchronized frequency $\omega$ increases, and the minimum value of the mass increases as well.

In the right panel of Fig.~\ref{excited-1s2p}, we observe that the minimal  mass  is always a constant value $M=1.31$ for the different azimuthal harmonic indexes  $m_2=1~,2,~3$, which is  similar to the $^1S^2S$ state case. Thus, three coordinates correspond to $(0.823,~1.31)$, $(0.871,~1.31)$ and  $(0.897,~1.31)$,  respectively.
In Table \ref{table4}, we present the  domain of existence of the  mass $\mu_2$ of the scalar field $\phi_2$ for the different values of $\Lambda_2$. We can see  that with the increase of coupling parameter $\Lambda_2$, the domain of existence of the $\mu_2$ begins to decrease.

\begin{table}[!htbp]
		\centering
		\begin{tabular}{|c|c|c|c|c|}
			\hline
\diagbox{$m_2$}{$\mu_{2}$}{$\Lambda_{2~~~}$}&$0$&$10$&$20$&$35$\\
            \hline
			$1$&$0.898\sim0.967$&$0.898\sim0.951$&$0.898\sim0.941$&$0.898\sim0.932$\\
			\hline
		\end{tabular}
\caption{The domain of existence of the  mass $\mu_2$ of the scalar field $\phi_2$ for different values of $\Lambda_2$ with the azimuthal harmonic index $m_2=1$. All solutions have $\mu_1=1$ and $m_1=1$.}
\label{table4}
	\end{table}
\subsection{ Case: The two coexisting states with self-interacting potential }
In the above two subsections, we gave two families of self-interacting multistate boson stars. Next, we consider the self-interacting potential $U(|\psi_1|)\neq0$, $U(|\psi_2|)\neq0$.
The matter Lagrangian ${\cal L}_{m}^{(3)}$ is given by
\begin{equation}
\label{lag}
 {\cal L}_{m}^{(3)}=-\nabla_a\psi_1^*\nabla^a\psi_1- U(|\psi_1|)-\nabla_a\psi_2^*\nabla^a\psi_2- U(|\psi_2|)\,,
\end{equation}
${\rm with}~U(|\psi_i|)= \mu_i^2\left|\psi_i\right|^2 + \lambda_i\left|\psi_i\right|^4$, we still use the same relation between the parameters $\Lambda_i$ and $\lambda_i$ given in Eq.(\ref{Lambda}), $i=1,2$.

\subsubsection{$^1S^2S$ state}
\begin{figure}[h!]
\begin{center}
\includegraphics[height=.27\textheight]{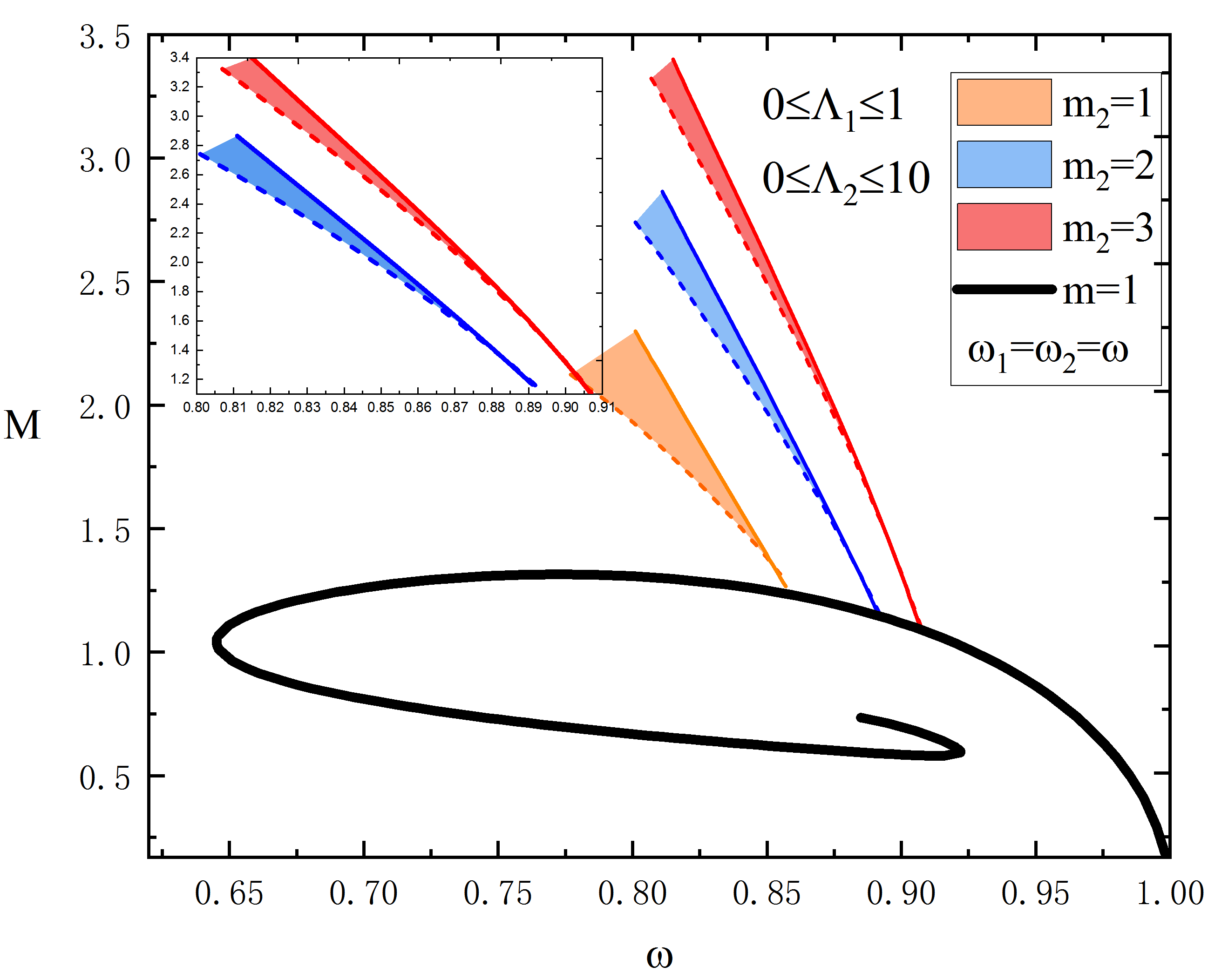}
\includegraphics[height=.27\textheight]{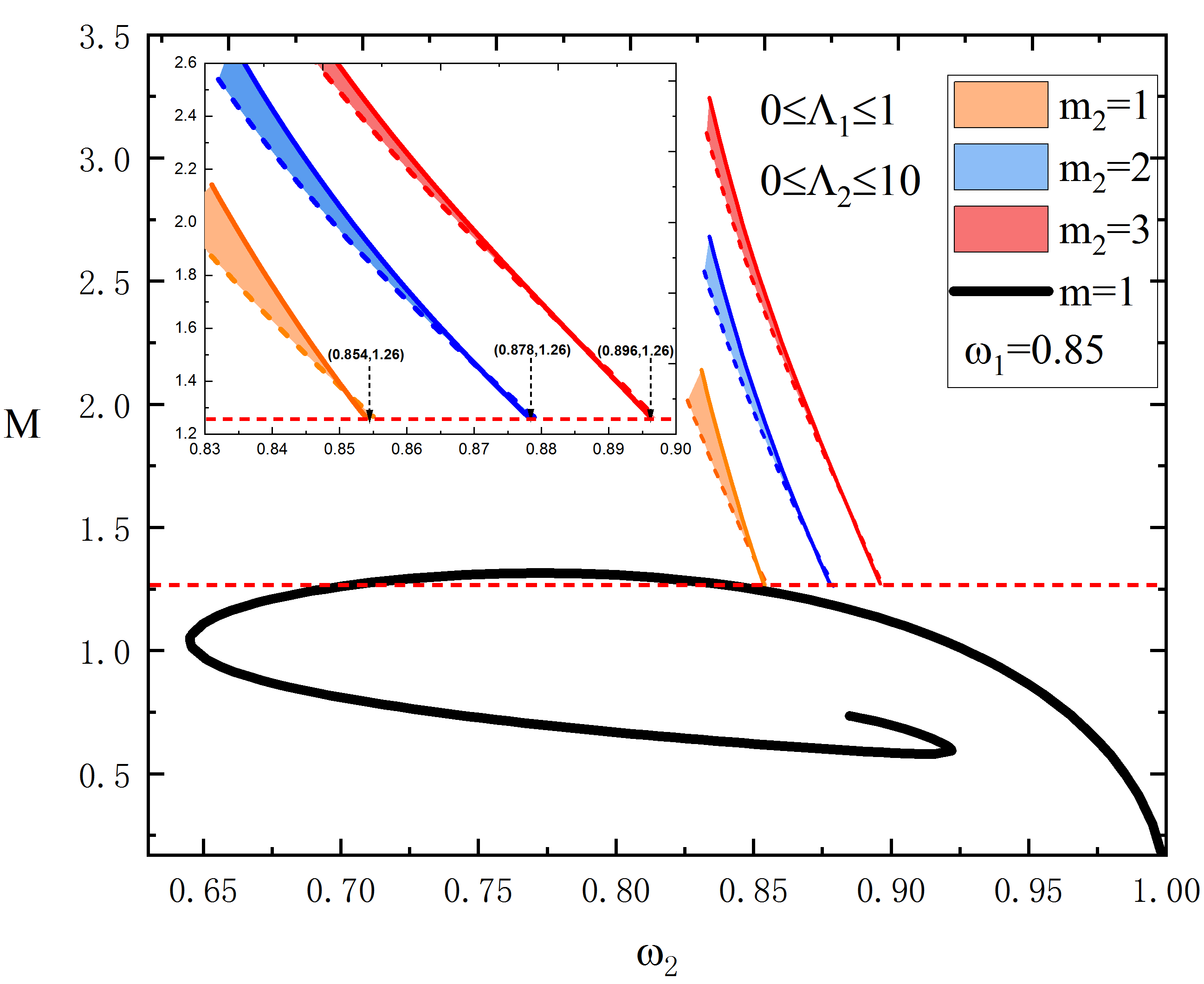}
\end{center}
\caption{\textit{Left}: The mass $M$ of the  SIMBSs as a function of the synchronized frequency $\omega$ with the azimuthal harmonic indexes $m_{2}=1~,2,~3$. \textit{Right}: The mass $M$ of the  SIMBSs as a function of the nonsynchronized frequency $\omega_2$ with $\omega_1=0.85$. The horizon red dashed line indicates the mass $M=1.26$. In both panels, the shaded regions corresponds to $0\leq\Lambda_1\leq1$ and $0\leq\Lambda_2\leq10$. The black spiral curve indicates the ground state with $m=1$ and all solutions have $\mu_1=1,\mu_2=0.93$, $m_1=1$.}
\label{coexistingstates-1s2s}
\end{figure}

According to the numerical results, we obtain the range of coupling parameters $\Lambda_1\in(0,2.8), \Lambda_2\in(0,388)$.
In the left panel of Fig.~\ref{coexistingstates-1s2s}, we exhibit the mass $M$ of the  SIMBSs versus the synchronized frequency $\omega$ for three sets of  SIMBSs with azimuthal harmonic index $m_{2}=1$ (the origin shaded region), $m_{2}=2$ (the blue shaded region), $m_{2}=3$ (the red shaded region), respectively. Here the shaded area corresponds to $0\leq\Lambda_1\leq1$ and $0\leq\Lambda_2\leq10$. We can see that the multistate boson stars curves have the similar behavior as the above two subsections case of $^1S^2S$ state. With the increase of the synchronized frequency $\omega$, the mass $M$ of the multistate boson stars decreases.

In the right panel of Fig.~\ref{coexistingstates-1s2s}, we show the mass $M$ of the multistate boson stars vs the nonsynchronized frequency $\omega_2$ with fixed value of $\omega_1=0.85$ for three sets of the azimuthal harmonic index. We found that the mass $M$ of the self-interacting multistate boson stars is larger than that of the free, multistate boson stars, under the common influence, which are both the coupling constants $\Lambda_1$  and  $\Lambda_2$.
Besides, in the inset of the right of Fig.~\ref{coexistingstates-1s2s}, we observe that the minimum value of mass of the SIMBSs is the constant value $M=1.26$, which is the same as $^1S^2S$ state of the first excited state with self-interacting potential. Therefore, three coordinates correspond to $(0.854,~1.26)$,  $(0.878,~1.26)$, and $(0.896,~1.26)$, respectively.

\begin{figure}[h!]
\begin{center}
\includegraphics[height=.27\textheight]{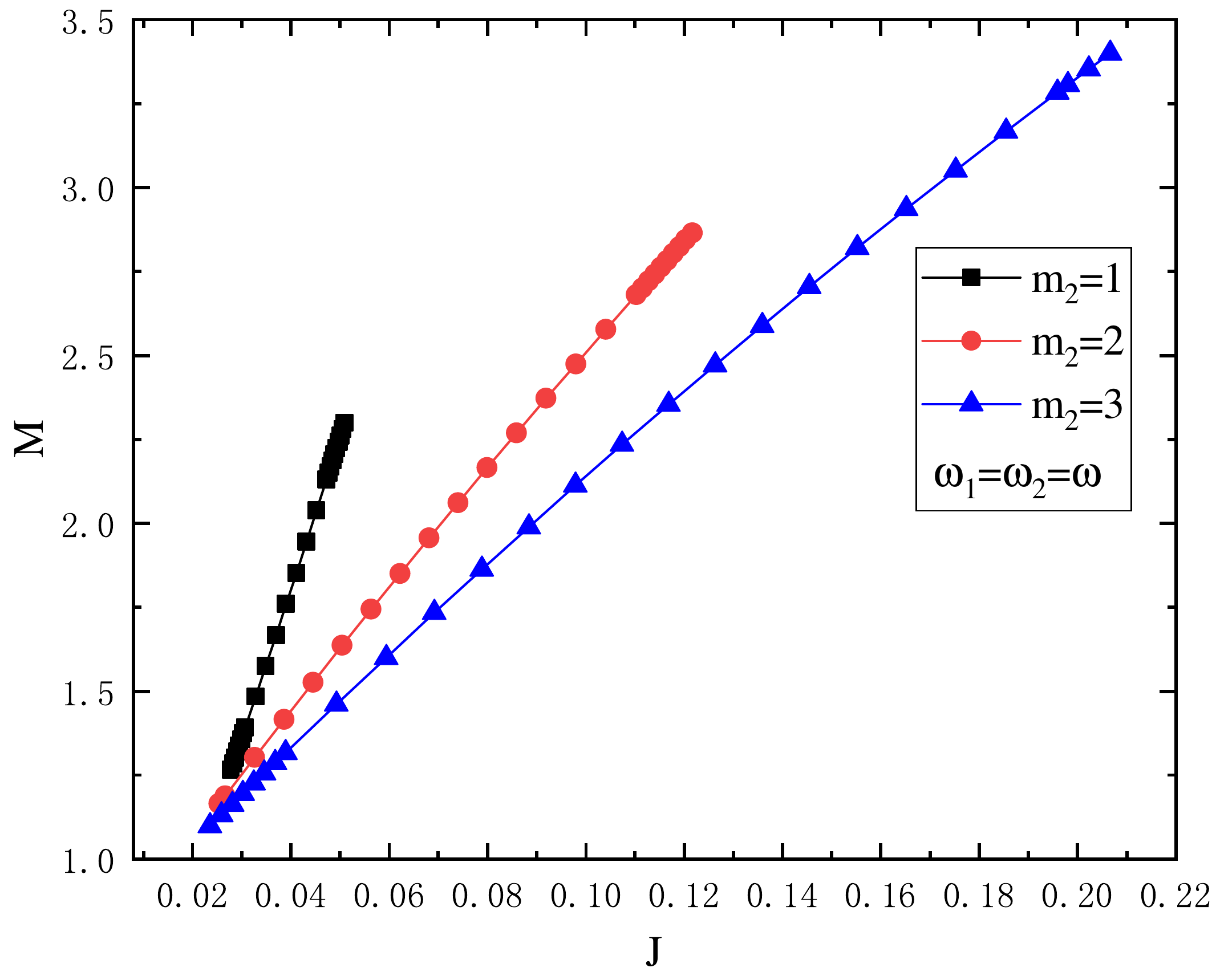}
\includegraphics[height=.267\textheight]{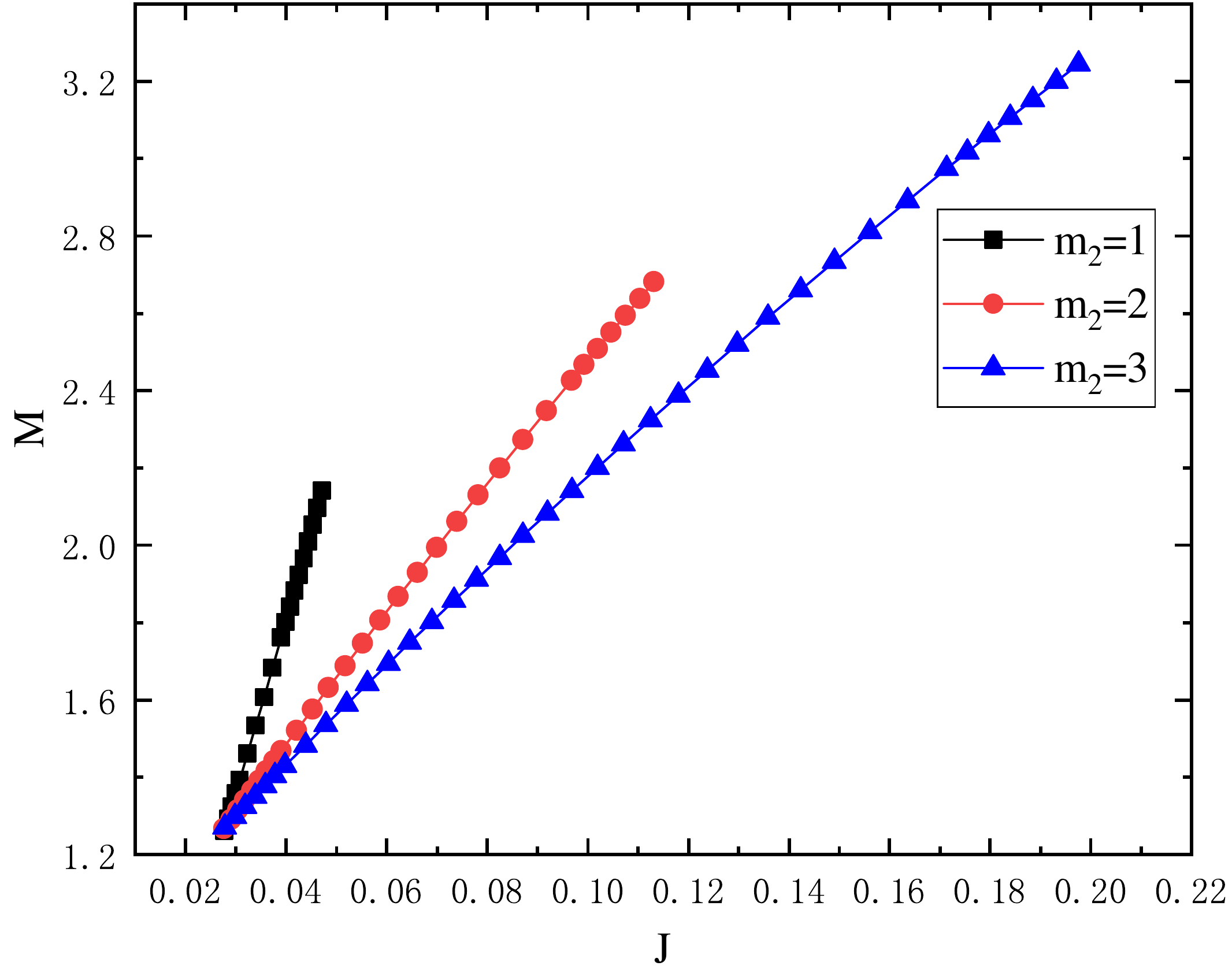}
\end{center}
\caption{\textit{Left}: The mass $M$ of  the SIMBSs  versus the angular momentum $J$ for the synchronized frequency $\omega$ with $m_{2}=1~,2,~3$, respectively. \textit{Right}: The mass $M$ of the SIMBSs versus the angular momentum $J$ for the nonsynchronized frequency $\omega_2$ with $m_{2}=1~,2,~3$, respectively, and we demand the parameter $\omega_1=0.85$.}
\label{coexistingstates-J}
\end{figure}
Next, we consider the variation of the solutions with the  mass  $M$ of the SIMBSs  that varies versus the angular momentum $J$, which is dependent on the frequency. In the left panel of Fig.~\ref{coexistingstates-J}, we exhibit the  mass $M$  of  the SIMBSs  versus the angular momentum $J$  with  different azimuthal harmonic index  $m_2 = 1$ (black lines),  2 (red lines), and  3 (blue lines) for the synchronized frequency $\omega$. Moreover, in the right panel of Fig.~\ref{coexistingstates-J}, the mass $M$  versus the angular momentum $J$  with  different azimuthal harmonic index $m_2=1~,2,~3$ for the nonsynchronized frequency $\omega_2$ are shown, and we set the frequency parameter $\omega_1=0.85$. Comparing with the results of the ground state boson stars in Ref.~\cite{Herdeiro:2015tia}, we can see that the case of the SIMBSs does not occur in zigzag patterns, and the minimum value of the mass $M$ is larger than that of the ground state boson stars.
\begin{table}[!htbp]
		\centering
		\begin{tabular}{|c|c|c|c|c|}
			\hline
\diagbox{$\Lambda_2$}{$\mu_{2}$}{$\Lambda_{1~~~}$}&$0$&$1$&$2$\\
            \hline
			$0$&$0.924\sim0.959$&$0.926\sim0.959$&$0.929\sim0.959$\\
			\hline
            $10$&$0.924\sim0.953$&$0.926\sim0.953$&$0.929\sim0.953$\\
            \hline
            $20$&$0.924\sim0.949$&$0.926\sim0.949$&$0.929\sim0.949$\\
            \hline
		\end{tabular}
\caption{The domain of existence of the  mass $\mu_2$ of the scalar field $\phi_2$ for different values of $\Lambda_1$ and $\Lambda_2$. All solutions have $\mu_1=1$, $m_1=1$ and $m_2=1$.}
\label{table5}
	\end{table}
In order to explore the influence of the different typical coupling parameters, we show  the domain of existence of $\mu_2$ for the different values of $\Lambda_1$ and $\Lambda_2$ in Table \ref{table5}. We observe that the range of mass $\mu_2$ decreases with the increase of the coupling parameters $\Lambda_1$ and $\Lambda_2$, which is similar to the case of $^1S^2S$ state in above two subsections.

\subsubsection{$^1S^2P$ state}
\begin{figure}[h!]
\begin{center}
\includegraphics[height=.25\textheight,width=.34\textheight, angle =0]{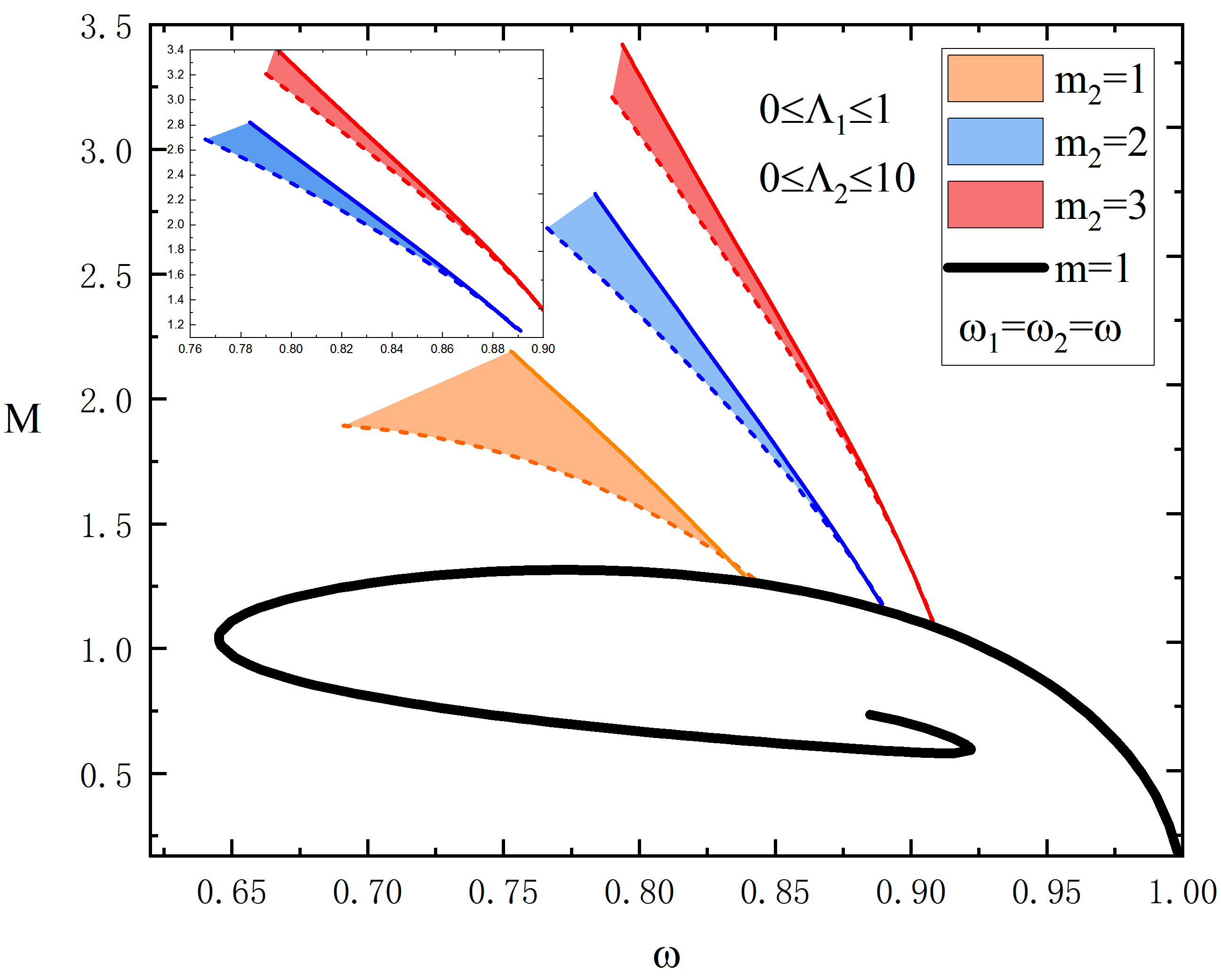}
\includegraphics[height=.25\textheight,width=.34\textheight, angle =0]{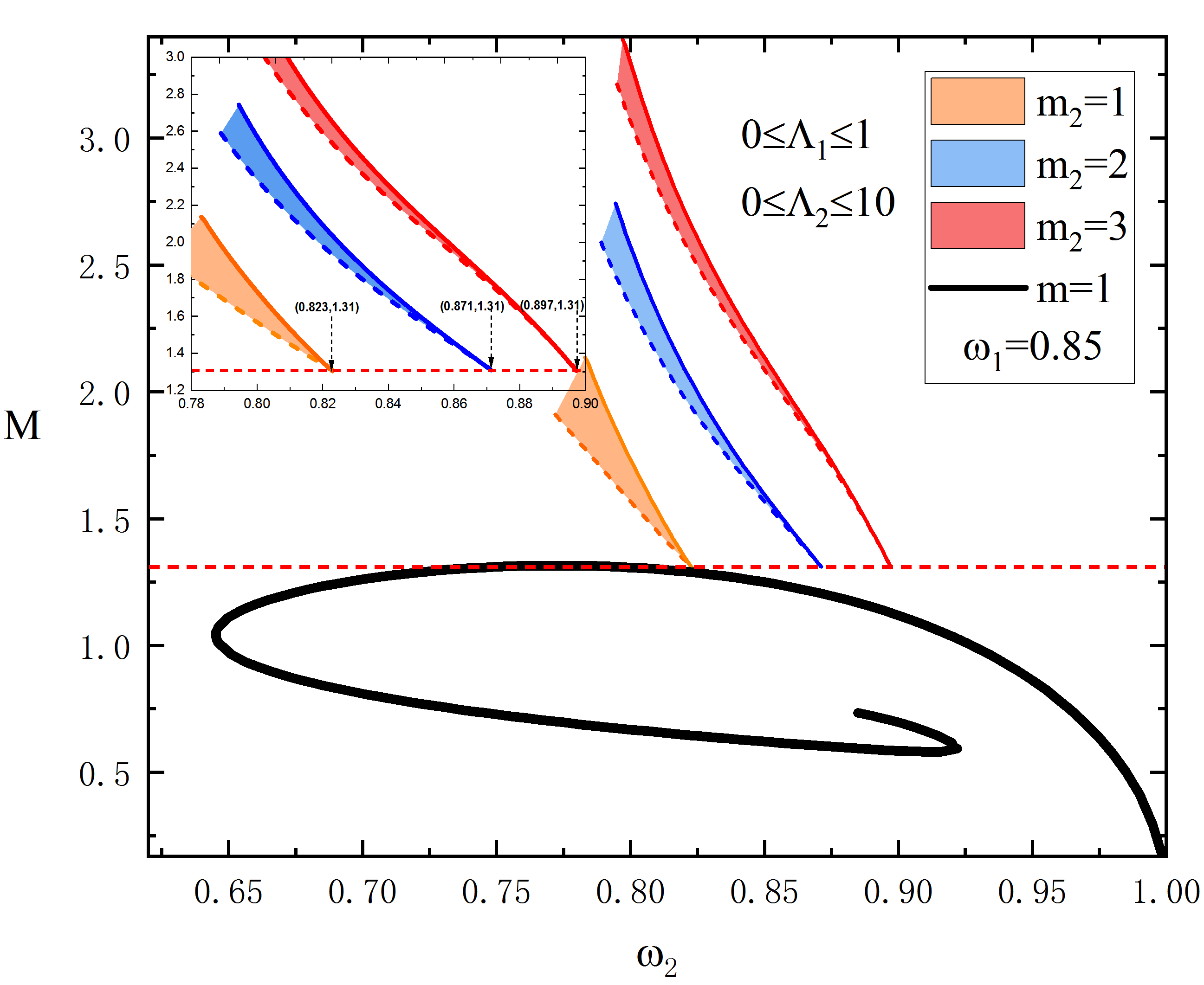}
\end{center}
\caption{\textit{Left}: The mass $M$ of the SIMBSs as a function of the synchronized frequency $\omega$ with the azimuthal harmonic indexes $m_{2}=1~,2,~3$. \textit{Right}: The mass $M$ of the  SIMBSs as a function of the nonsynchronized frequency $\omega_2$ with $\omega_1=0.85$. The horizon red dashed line indicates the mass $M=1.31$. In both panels, the shaded regions corresponds to $0\leq\Lambda_1\leq1$ and $0\leq\Lambda_2\leq10$. The black spiral curve indicates the ground state with $m=1$ and all solutions have $\mu_1=1,\mu_2=0.93$, $m_1=1$.}
\label{coexistingstates-1s2p}
\end{figure}

According to the numerical results, we obtain the range of coupling parameters $\Lambda_1\in(0,5.8), \Lambda_2\in(0,41)$.
In the left panel of Fig.~\ref{coexistingstates-1s2p}, we plot the mass $M$ of the SIMBSs vs the synchronized frequency $\omega$ with the azimuthal harmonic index $m_{2}=1$ (the origin area), $m_{2}=2$ (the blue area), $m_{2}=3$ (the red area), respectively.
In the inset of the left panel of Fig.~\ref{coexistingstates-1s2p},  we show the detail of the blue and red  shaded regions with  $m_{2}=2$ and $m_{2}=3$.
From the graphic, we found that the SIMBSs  exists only a stable branch, which is similar to the $^1S^2P$ state in above two subsections.
Moreover, we found that the mass of multistate boson stars decreases with increasing the frequency $\omega$.

\begin{table}[!htbp]
		\centering
		\begin{tabular}{|c|c|c|c|c|}
			\hline
\diagbox{$\Lambda_2$}{$\mu_{2}$}{$\Lambda_{1~~~}$}&$0$&$1$&$2$\\
            \hline
			$0$&$0.898\sim0.967$&$0.904\sim0.967$&$0.909\sim0.967$\\
			\hline
            $10$&$0.898\sim0.951$&$0.904\sim0.951$&$0.909\sim0.951$\\
            \hline
            $20$&$0.898\sim0.941$&$0.904\sim0.941$&$0.909\sim0.941$\\
            \hline
		\end{tabular}
\caption{The domain of existence of the  mass $\mu_2$ of the scalar field $\phi_2$ for  different values of $\Lambda_1$ and $\Lambda_2$. All solutions have $\mu_1=1$, $m_1=1$ and $m_2=1$.}
\label{table6}
	\end{table}
In the right panel of Fig.~\ref{coexistingstates-1s2p}, we observe that the minimal  mass  is always a constant value $M=1.31$ for the different azimuthal harmonic indexes  $m_2=1~,2,~3$, which is  similar to the $^1S^2P$ state of the first excited state with self-interacting potential case. Therefore, three coordinates correspond to $(0.823,~1.31)$, $(0.871,~1.31)$ and  $(0.897,~1.31)$,  respectively.
To explore the influence of the different typical coupling parameters, we exhibit for various values of $\Lambda_1$ and $\Lambda_2$, the domain of existence of mass $\mu_2$ is shown in Table \ref{table6}. We observe that the range of mass $\mu_2$ decreases with the increase of the coupling parameters $\Lambda_1$ and $\Lambda_2$, which is similar to the case of $^1S^2S$ state of the two coexisting states with self-interacting potential case.

\section{Conclusion}\label{sec5}

In this paper, we  have constructed  rotating boson stars composed of the coexisting states of two quartic-self-interacting scalar fields,
including the ground and first excited states. Comparing with the solutions of the quartic-BSs in Ref.~\cite{Herdeiro:2015tia},
we have found that the SIMBSs have two types of nodes, including the $^1S^2S$ and $^1S^2P$  states.
According to the numerical results, we find that the coupling parameter $\Lambda$ is the finite value for the $^1S^2S$  and   $^1S^2P$  states, which is different from the quartic-BSs in Ref.~\cite{Herdeiro:2015tia}.
By calculating the coexisting phase of the SIMBSs, we found that the domain of existence of the mass $\mu_2$ decreases with increasing the coupling parameters.
When the synchronized frequency $\omega$ tends to its maximum,  the first excited state could reduce to zero,  and there exists only the ground state. However, for the case of nonsynchronized frequency $\omega_2$, we fixed the value of $\omega_1=0.85$, the minimal mass of the SIMBSs is always a constant value for the different azimuthal harmonic indexes $m_2=2,~3$.

It is important to understand how the stability properties of the self-interacting multistate boson stars, according to numerical analysis of  the stability of the self-interaction boson stars in~\cite{Kleihaus:2011sx}, the authors found that the solutions have both stable and unstable branches. Besides, in 2010,
Bernal {\em et al.}~\cite{Bernal:2009zy} found for the nonrotating multistate boson stars solutions, such configurations are stable if the number of particles in the ground
state is larger than the number of particles in the excited state.
To our knowledge, it is difficult to numerically analyze the stability properties of self-interaction multistate boson stars.
However, a good way to guarantee the stability is to have the scalar field perturbations decouple from the metric perturbations in Ref.~\cite{Ganchev:2017uuo}.

There are several interesting extensions of our work.
Firstly, we have studied the SIMBSs, next, we will investigate the gravitational redshift and the RC of  multistate boson stars inspired by the work~\cite{Schunck:1997dn,Deng:1998}.
In 2019, Brihaye and Hartmann have studied the spontaneous scalarization of spherically symmetric, asymptotically flat boson stars in Ref.~\cite{Brihaye:2019puo}.
Therefore, the second extension of our work is to study the spontaneous scalarization for axially symmetric BSs solutions.
Finally, we are planning to construct this solutions where the self-interaction potential of the scalar field $\psi$ coupled minimally to Einstein gravity in higher-dimensional spacetime.

\section*{Acknowledgement}
Hong-Bo Li would like to thank  Tong-Tong Hu  and Shuo Sun  for  helpful discussion.
Some computations were performed on the   Shared Memory system at  Institute of Computational Physics and Complex Systems in Lanzhou University.

\providecommand{\href}[2]{#2}\begingroup\raggedright
\endgroup

\end{document}